\documentclass[11pt]{article}
\usepackage[margin=1in]{geometry}
\usepackage{amsmath,amssymb,amsthm}
\usepackage{mathtools}
\usepackage{bm}
\usepackage{graphicx}
\usepackage{hyperref}
\usepackage{enumitem}
\usepackage{algorithm}
\usepackage{algpseudocode}

\newcommand{\vt}{\boldsymbol{\vartheta}}
\newcommand{\bx}{\mathbf{x}}
\newcommand{\bp}{\mathbf{p}}
\newcommand{\by}{\mathbf{y}}
\newcommand{\bz}{\mathbf{z}}

\newcommand{\bw}{\mathbf{w}}
\newcommand{\R}{\mathbb{R}}
\newcommand{\E}{\mathbb{E}}
\newcommand{\ii}{\mathrm{i}}

\newtheorem{remark}{Remark}

\begin{document}

\begin{center}
{\LARGE\bfseries Weak Adversarial Neural Pushforward Method \\[4pt]
for the Wigner--Moyal Equation}

\bigskip

{\large Andrew Qing He\textsuperscript{*} \quad Wei Cai\textsuperscript{*}
\quad Sihong Shao\textsuperscript{\dag}}

\medskip

{\itshape\textsuperscript{*}Department of Mathematics, Southern Methodist
University, Dallas, TX, USA.}\\
{\ttfamily andrewho@smu.edu, cai@smu.edu}

\medskip

{\itshape\textsuperscript{\dag}School of Mathematical Sciences, Peking
University, Beijing, China.}\\
{\ttfamily sihong@math.pku.edu.cn}

\bigskip

\today
\end{center}

\bigskip

\begin{abstract}
We extend the Weak Adversarial Neural Pushforward Method to the Wigner--Moyal
equation governing the phase-space dynamics of quantum systems.  The central
contribution is a structural observation: integrating the nonlocal
pseudo-differential potential operator against plane-wave test functions
produces a Dirac delta that exactly inverts the Fourier transform defining the
Wigner potential kernel, reducing the operator to a pointwise finite difference
of the potential at two shifted arguments.  This holds in arbitrary dimension,
requires no truncation of the Moyal series, and treats the potential as a
black-box function oracle with no derivative information.  To handle the
negativity of the Wigner quasi-probability distribution, we introduce a signed
pushforward architecture that decomposes the solution into two non-negative
phase-space distributions mixed with a learnable weight.  The resulting method
inherits the mesh-free, Jacobian-free, and scalable properties of the original
framework while extending it to the quantum setting.  Numerical experiments
on the Wigner--Fokker--Planck equation with both harmonic and anharmonic
potentials validate the approach against analytical and spectral reference
solutions.
\end{abstract}

\medskip
\noindent\textbf{Keywords:}
Wigner--Moyal equation, Wigner function, quantum phase space,
neural pushforward map, weak adversarial network,
plane-wave test functions, pseudo-differential operator

\medskip
\noindent\textbf{MSC 2020:} 65N75, 68T07, 81Q05, 81S30

\section{Introduction}\label{sec:intro}

The Wigner function, introduced by Wigner in 1932~\cite{Wigner1932}, provides
a phase-space representation of quantum mechanics that parallels the classical
Liouville description.  The evolution of the Wigner function $f(t,\bx,\bp)$ is
governed by the Wigner--Moyal equation~\cite{Moyal1949}, which differs from the
classical Liouville equation by a nonlocal pseudo-differential potential
operator $\Theta[V]$.  This operator is defined through a double integral
involving the Fourier transform of the antisymmetrized potential, and its
nonlocality is the mathematical expression of quantum interference in phase
space.

Solving the Wigner--Moyal equation numerically is challenging for several
reasons.  First, the equation lives in a $2N$-dimensional phase space, where
$N$ is the number of spatial degrees of freedom, making grid-based methods
prohibitively expensive for all but the lowest-dimensional problems.  Second,
the pseudo-differential operator $\Theta[V]$ is nonlocal in the momentum
variable, requiring either its explicit evaluation (which involves a Fourier
transform at each grid point) or a truncation of the Moyal series to a finite
number of momentum derivatives.  The latter produces the truncated Wigner
approximation (TWA)~\cite{Steel1998,Polkovnikov2010}, which is widely used in
quantum optics and cold-atom physics but introduces systematic errors that grow
with $\hbar$ and with the degree of anharmonicity of the potential.  Third, the
Wigner function is a quasi-probability distribution that can take negative
values, precluding the direct use of probabilistic particle methods that rely
on non-negative densities.

Stochastic approaches to the Wigner equation have been developed based on the
signed-particle formulation~\cite{Nedjalkov2004,Sellier2015}, in which
particles carry positive or negative weights and undergo branching processes
that encode the quantum potential.  While these methods avoid spatial
discretization, they suffer from the numerical sign problem: the variance of
estimators grows exponentially with time as the number of sign changes
accumulates.

Neural network methods for solving partial differential equations have seen
rapid development in recent years, including Physics-Informed Neural Networks
(PINNs)~\cite{PINN}, the Deep Ritz method~\cite{deepritz}, and Weak
Adversarial Networks (WAN)~\cite{wan}.  In~\cite{He2025}, the Weak Adversarial
Neural Pushforward Method (WANPM) was introduced for solving Fokker--Planck
equations.  WANPM learns a neural pushforward map that transforms samples from
a simple base distribution into samples from the solution distribution, with
training guided by a weak formulation employing computationally efficient
plane-wave test functions.  The method has since been extended to
McKean--Vlasov equations~\cite{HeCai2026MV} and to Fokker--Planck equations on
Riemannian manifolds~\cite{HeCai2026manifold}.

The present paper extends WANPM to the Wigner--Moyal equation.  The key finding
is a structural compatibility between the plane-wave test functions used in
WANPM and the Fourier-integral definition of the pseudo-differential operator
$\Theta[V]$: when the weak-form integral $\int(\Theta[V]f)\,\varphi\,
d\bx\,d\bp$ is evaluated with a plane-wave test function $\varphi =
\sin(\bw_x\cdot\bx + \bw_p\cdot\bp + \kappa t + b)$, the Fourier kernel in
$\Theta[V]$ combines with the plane wave to produce a Dirac delta that
collapses the integral, yielding an expectation over $f$ involving only $V$
evaluated at two shifted positions $\bx\pm\frac{\hbar}{2}\bw_p$.  This
reduction is exact --- it requires no truncation of the Moyal series --- and
treats $V$ as a black-box function oracle, requiring no derivative information.

The paper is organized as follows.  Section~\ref{sec:wigner} introduces the
Wigner--Moyal equation and the pseudo-differential operator $\Theta[V]$.
Section~\ref{sec:weak} derives the weak formulation and evaluates the potential
integral, establishing the central result.  Section~\ref{sec:residual}
assembles the complete weak-form residual and discusses the connection to the
classical Liouville equation and the truncated Wigner approximation.
Section~\ref{sec:signed} introduces the signed pushforward architecture for
handling Wigner negativity.  Section~\ref{sec:algorithm} describes the
adversarial training algorithm.  Section~\ref{sec:numerics} presents
numerical experiments on the Wigner--Fokker--Planck equation.
Section~\ref{sec:conclusion} concludes.

\section{The Wigner--Moyal Equation}\label{sec:wigner}

Consider a quantum system with $N$ degrees of freedom, positions
$\bx = (x_1,\ldots,x_N)\in\R^N$, momenta $\bp = (p_1,\ldots,p_N)\in\R^N$,
mass $m$, and potential $V(\bx)$.  The Wigner function
$f(t,\bx,\bp)$~\cite{Wigner1932} evolves according to the Wigner--Moyal
equation~\cite{Moyal1949}:
\begin{equation}\label{eq:wigner}
  \frac{\partial f}{\partial t}
  + \sum_{i=1}^N \frac{p_i}{m}\frac{\partial f}{\partial x_i}
  = \Theta[V]\,f,
\end{equation}
where the pseudo-differential operator $\Theta[V]$ is defined by
\begin{equation}\label{eq:theta}
  (\Theta[V]\,f)(t,\bx,\bp)
  = \frac{1}{\ii\hbar}\frac{1}{(2\pi\hbar)^N}
    \int_{\R^N}\!\!\int_{\R^N}
    \left[
      V\!\left(\bx+\tfrac{\by}{2}\right)
      - V\!\left(\bx-\tfrac{\by}{2}\right)
    \right]
    \mathrm{e}^{\ii(\bp-\bp')\cdot\by/\hbar}\,
    f(t,\bx,\bp')\;\mathrm{d}\by\,\mathrm{d}\bp'.
\end{equation}
The left-hand side of~\eqref{eq:wigner} is the classical Liouville operator
(free streaming along Hamiltonian trajectories); the right-hand side encodes
the quantum potential, which is nonlocal in the momentum variable.

The Wigner function satisfies the normalization
$\int f\,\mathrm{d}\bx\,\mathrm{d}\bp = 1$ and the marginal conditions
\begin{equation}\label{eq:marginals}
  \int_{\R^N} f(t,\bx,\bp)\,\mathrm{d}\bp = |\psi(t,\bx)|^2 \geq 0,
  \qquad
  \int_{\R^N} f(t,\bx,\bp)\,\mathrm{d}\bx = |\hat\psi(t,\bp)|^2 \geq 0,
\end{equation}
where $\psi$ and $\hat\psi$ are the position- and momentum-space wave
functions.  However, $f$ itself can take negative values; it is a
quasi-probability distribution.

\section{Weak Formulation and the Central Result}\label{sec:weak}

\subsection{Setup}

Multiply~\eqref{eq:wigner} by the plane-wave test function
\begin{equation}\label{eq:test}
  \varphi^{(k)}(t,\bx,\bp)
  = \sin(\bw_x^{(k)}\!\cdot\bx + \bw_p^{(k)}\!\cdot\bp
    + \kappa^{(k)} t + b^{(k)}),
\end{equation}
with trainable parameters $\bw_x^{(k)},\bw_p^{(k)}\in\R^N$,
$\kappa^{(k)},b^{(k)}\in\R$, and integrate over
$(\bx,\bp)\in\R^{2N}$ and $t\in[0,T]$.  Integrating by parts in $t$ and in
$\bx$ (for the transport term on the left-hand side of~\eqref{eq:wigner}), we
obtain
\begin{multline}\label{eq:weak_form}
  \int_{\R^{2N}}\! f\,\varphi^{(k)}\Big|_{t=T}\mathrm{d}\bx\,\mathrm{d}\bp
  \;-\;
  \int_{\R^{2N}}\! f\,\varphi^{(k)}\Big|_{t=0}\mathrm{d}\bx\,\mathrm{d}\bp \\
  -\;
  \int_0^T\!\!\int_{\R^{2N}}\! f\left(
    \frac{\partial\varphi^{(k)}}{\partial t}
    + \sum_i\frac{p_i}{m}\frac{\partial\varphi^{(k)}}{\partial x_i}
  \right)\mathrm{d}\bx\,\mathrm{d}\bp\,\mathrm{d}t
  \;=\;
  \int_0^T I^{(k)}\;\mathrm{d}t,
\end{multline}
where the right-hand side contains the potential integral
\begin{equation}\label{eq:I_def}
  I^{(k)} = \int_{\R^{2N}} (\Theta[V]\,f)\;\varphi^{(k)}\;\mathrm{d}\bx\,\mathrm{d}\bp.
\end{equation}
The left-hand side is standard: the transport term produces the pointwise
expression $[\kappa^{(k)} + \bw_x^{(k)}\cdot\bp/m]\cos\phi^{(k)}$, where
$\phi^{(k)} = \bw_x^{(k)}\cdot\bx + \bw_p^{(k)}\cdot\bp + \kappa^{(k)} t
+ b^{(k)}$.  The central task is to evaluate $I^{(k)}$.

\subsection{Evaluation of the potential integral $I^{(k)}$}

Substituting~\eqref{eq:theta} and~\eqref{eq:test} into~\eqref{eq:I_def}
and writing $\alpha := \bw_x^{(k)}\cdot\bx + \kappa^{(k)} t + b^{(k)}$:
\begin{equation}\label{eq:I_expanded}
  I^{(k)} = \frac{1}{\ii\hbar}\frac{1}{(2\pi\hbar)^N}
  \int_{\R^{4N}}
  \left[
    V\!\left(\bx+\tfrac{\by}{2}\right)
    - V\!\left(\bx-\tfrac{\by}{2}\right)
  \right]
  \mathrm{e}^{\ii(\bp-\bp')\cdot\by/\hbar}\,
  f(t,\bx,\bp')\,
  \sin(\alpha + \bw_p^{(k)}\!\cdot\bp)
  \;\mathrm{d}\by\,\mathrm{d}\bp'\,\mathrm{d}\bp\,\mathrm{d}\bx.
\end{equation}

\paragraph{Step 1: $\bp$-integration.}
Writing
$\sin(\alpha+\bw_p^{(k)}\cdot\bp)
= \frac{1}{2\ii}(\mathrm{e}^{\ii(\alpha+\bw_p^{(k)}\cdot\bp)}
- \mathrm{e}^{-\ii(\alpha+\bw_p^{(k)}\cdot\bp)})$, the $\bp$-dependent
factors are
$\mathrm{e}^{\ii\bp\cdot(\by/\hbar\pm\bw_p^{(k)})}$, and the $\bp$-integral
produces
\begin{equation}\label{eq:delta}
  \int_{\R^N} \mathrm{e}^{\ii\bp\cdot(\by/\hbar\pm\bw_p^{(k)})}\;\mathrm{d}\bp
  = (2\pi\hbar)^N\,\delta^{(N)}(\by\pm\hbar\bw_p^{(k)}).
\end{equation}

\paragraph{Step 2: $\by$-integration.}
Define
$D(\bx) := V(\bx+\frac{\hbar}{2}\bw_p^{(k)})
         - V(\bx-\frac{\hbar}{2}\bw_p^{(k)})$.
The delta functions set $\by = -\hbar\bw_p^{(k)}$ (Branch~$+$, giving
potential difference $-D$, exponential factor
$\mathrm{e}^{\ii\bw_p^{(k)}\cdot\bp'}$) and $\by = +\hbar\bw_p^{(k)}$
(Branch~$-$, giving $+D$, factor $\mathrm{e}^{-\ii\bw_p^{(k)}\cdot\bp'}$).
Including the $\frac{\pm 1}{2\ii}\mathrm{e}^{\pm\ii\alpha}$ prefactors from
the sine decomposition:
\begin{align}
  \text{Branch $+$:}\quad
    &-\frac{(2\pi\hbar)^N}{2\ii}\,D\,f\,
    \mathrm{e}^{\ii(\alpha+\bw_p^{(k)}\cdot\bp')},
  \label{eq:br_plus}\\
  \text{Branch $-$:}\quad
    &-\frac{(2\pi\hbar)^N}{2\ii}\,D\,f\,
    \mathrm{e}^{-\ii(\alpha+\bw_p^{(k)}\cdot\bp')}.
  \label{eq:br_minus}
\end{align}
Summing:
\begin{equation}\label{eq:sum_branches}
  -\frac{(2\pi\hbar)^N}{\ii}\,D(\bx)\,f(t,\bx,\bp')\,
  \cos(\bw_x^{(k)}\!\cdot\bx+\bw_p^{(k)}\!\cdot\bp'+\kappa^{(k)} t+b^{(k)}).
\end{equation}

\paragraph{Step 3: prefactor and $(\bx,\bp')$-integration.}
Multiplying~\eqref{eq:sum_branches} by the overall prefactor
$\frac{1}{\ii\hbar}\frac{1}{(2\pi\hbar)^N}$ and using
$\frac{-1}{\ii^2} = 1$:
\begin{equation}\label{eq:I_result}
  I^{(k)} = \E_{(\bx,\bp)\sim f}\!\left[
    \frac{V\!\left(\bx+\frac{\hbar\bw_p^{(k)}}{2}\right)
          - V\!\left(\bx-\frac{\hbar\bw_p^{(k)}}{2}\right)}{\hbar}\;
    \cos\phi^{(k)}
  \right],
\end{equation}
where we renamed $\bp'\to\bp$.

\begin{remark}
The nonlocal pseudo-differential operator $\Theta[V]$ has been reduced to a
pointwise function of $(\bx,\bp)$ inside an expectation over~$f$.  This
reduction is exact and holds for any potential $V$ and any dimension~$N$.  The
mechanism is that the plane-wave test function produces a Dirac delta
via~\eqref{eq:delta} that exactly inverts the Fourier transform defining the
Wigner potential kernel.
\end{remark}

\section{The Complete Weak-Form Residual}\label{sec:residual}

Combining the transport contribution from the left-hand side
of~\eqref{eq:weak_form} with the potential integral~\eqref{eq:I_result}, the
weak-form residual for each test function $\varphi^{(k)}$ is:
\begin{equation}\label{eq:residual}
\begin{aligned}
  R^{(k)} &= \E_{f(T,\cdot)}\!\left[\varphi^{(k)}(T,\cdot)\right]
            - \E_{f_0}\!\left[\varphi^{(k)}(0,\cdot)\right] \\[4pt]
           &\quad- \int_0^T \E_{(\bx,\bp)\sim f(t,\cdot)}\!\Bigg[
             \bigg(
               \kappa^{(k)}
               + \frac{\bw_x^{(k)}\!\cdot\bp}{m} \\
           &\qquad\qquad
               - \frac{V\!\left(\bx+\frac{\hbar\bw_p^{(k)}}{2}\right)
                       - V\!\left(\bx-\frac{\hbar\bw_p^{(k)}}{2}\right)}{\hbar}
             \bigg)
             \cos\phi^{(k)}
           \Bigg] \mathrm{d}t.
\end{aligned}
\end{equation}
The integrand is a pointwise function of $(\bx,\bp)$, despite the original
operator $\Theta[V]$ being nonlocal, and is therefore directly estimable by
Monte Carlo over pushforward samples.

\begin{remark}
For the truncated Wigner equation, the operator $\Theta[V]$ is approximated by
a differential operator (retaining only the $V'$ and $V'''$ terms), and one can
write a pointwise backward operator $\mathcal{L}\varphi$.  For the full
Wigner--Moyal equation, such a pointwise operator does not exist: the reduction
to a pointwise expression holds only after taking the expectation over~$f$.
The plane-wave test function structure is essential for this reduction.
\end{remark}

\subsection{Classical limit}

When $\hbar\to 0$:
\begin{equation}\label{eq:classical_limit}
  \frac{V(\bx+\frac{\hbar}{2}\bw_p) - V(\bx-\frac{\hbar}{2}\bw_p)}{\hbar}
  \;\longrightarrow\;
  \nabla V(\bx)\cdot\bw_p,
\end{equation}
and~\eqref{eq:residual} reduces to the weak-form residual for the classical
Liouville equation, with the integrand
$[\kappa + \bw_x\cdot\bp/m - \nabla V\cdot\bw_p]\cos\phi$.

\subsection{Recovery of the truncated Wigner equation}

Taylor-expanding with $\boldsymbol{\delta} := \frac{\hbar}{2}\bw_p$:
\begin{multline}\label{eq:taylor}
  \frac{V(\bx+\boldsymbol{\delta}) - V(\bx-\boldsymbol{\delta})}{\hbar}
  = \nabla V\!\cdot\bw_p
  + \frac{\hbar^2}{24}\sum_{i,j,k}
    V_{x_i x_j x_k}\,w_{p,i}\,w_{p,j}\,w_{p,k} \\
  + \frac{\hbar^4}{1920}\sum_{i,j,k,l,m}
    V_{x_i x_j x_k x_l x_m}\,w_{p,i}\,w_{p,j}\,w_{p,k}\,w_{p,l}\,w_{p,m}
  + \cdots
\end{multline}
Only odd-order derivatives survive (the even terms cancel by antisymmetry).
The first line gives the classical Liouville operator; the first two terms give
the truncated Wigner equation; the full expression~\eqref{eq:residual} resums
the entire series into a single finite difference.

\section{Signed Pushforward Architecture}\label{sec:signed}

Unlike the Fokker--Planck or McKean--Vlasov settings, the Wigner function can
take negative values.  We handle this by decomposing $f$ into positive and
negative parts, each represented by its own pushforward network.

\subsection{Decomposition}

We write
\begin{equation}\label{eq:decomposition}
  f(t,\bx,\bp)
  = \alpha^+\,f^+(t,\bx,\bp) - \alpha^-\,f^-(t,\bx,\bp),
\end{equation}
where $f^\pm\geq 0$ are non-negative densities on $\R^{2N}$, each normalized
to~$1$, and the constants satisfy
\begin{equation}\label{eq:alpha_constraint}
  \alpha^+ - \alpha^- = 1,
\end{equation}
guaranteeing $\int f\,\mathrm{d}\bx\,\mathrm{d}\bp = 1$.  At $t=0$, the
decomposition $f_0 = \alpha^+ f_0^+ - \alpha^- f_0^-$ is prescribed and we
know how to sample from $f_0^\pm$ independently.

There are infinitely many valid decompositions of a given $f$ into the
form~\eqref{eq:decomposition}; the adversarial training selects the one that
best satisfies the weak formulation.

\subsection{Neural pushforward networks}

We introduce two independent neural networks with independent parameter sets:
\begin{equation}\label{eq:networks}
  F^+_{\vt^+}:\R\times\R^{2N}\times\R^{d^+_{\mathrm{base}}} \to \R^{2N},
  \qquad
  F^-_{\vt^-}:\R\times\R^{2N}\times\R^{d^-_{\mathrm{base}}} \to \R^{2N}.
\end{equation}
Each network maps from time $t$, an initial phase-space point
$(\bx_0,\bp_0)\in\R^{2N}$, and base noise $\bz\in\R^{d_{\mathrm{base}}}$ to a
phase-space point $(\bx,\bp)\in\R^{2N}$.  For each sample~$m$:
\begin{enumerate}[label=(\roman*)]
  \item Draw $(\bx_0^{+,(m)}, \bp_0^{+,(m)}) \sim f_0^+$ and
    $(\bx_0^{-,(m)}, \bp_0^{-,(m)}) \sim f_0^-$ independently.
  \item Draw $\bz^{+,(m)}\sim\pi^+_{\mathrm{base}}$ and
    $\bz^{-,(m)}\sim\pi^-_{\mathrm{base}}$ independently.
  \item Compute:
    \begin{align}\label{eq:samples}
      (\bx^{+,(m)}, \bp^{+,(m)})
        &= F^+_{\vt^+}(t, \bx_0^{+,(m)}, \bp_0^{+,(m)}, \bz^{+,(m)}),
      \notag\\
      (\bx^{-,(m)}, \bp^{-,(m)})
        &= F^-_{\vt^-}(t, \bx_0^{-,(m)}, \bp_0^{-,(m)}, \bz^{-,(m)}).
    \end{align}
\end{enumerate}

\subsection{Initial condition enforcement}

To enforce the initial condition exactly at $t=0$:
\begin{equation}\label{eq:architecture}
  F^\pm_{\vt^\pm}(t, \bx_0, \bp_0, \bz)
  = (\bx_0, \bp_0) + \sqrt{t}\;
    \widetilde{F}^\pm_{\vt^\pm}(t, \bx_0, \bp_0, \bz),
\end{equation}
where $\widetilde{F}^\pm_{\vt^\pm}$ are unconstrained neural networks.  At
$t=0$, $F^\pm(0,\bx_0,\bp_0,\bz) = (\bx_0,\bp_0)$, recovering $f_0^\pm$
exactly.

\subsection{Learnable mixing weight}

We parameterize $\alpha^+ = 1 + \alpha$, $\alpha^- = \alpha$ with a single
trainable scalar $\alpha\geq 0$, enforcing~\eqref{eq:alpha_constraint} by
construction.  At initialization, $\alpha = 0$ if $f_0\geq 0$ (e.g., a
coherent state); otherwise, $\alpha$ is initialized from the negative volume
of~$f_0$.  The parameter $\alpha$ is trained jointly with the network
parameters; the initial condition is enforced by construction, not by penalty.

\subsection{Monte Carlo estimator}

For any function $\varphi(\bx,\bp)$:
\begin{equation}\label{eq:mc_estimator}
  \int_{\R^{2N}} \varphi\,f\;\mathrm{d}\bx\,\mathrm{d}\bp
  \;\approx\;
  \frac{1}{M}\sum_{m=1}^M
  \left[
    \alpha^+\,\varphi(\bx^{+,(m)}, \bp^{+,(m)})
    - \alpha^-\,\varphi(\bx^{-,(m)}, \bp^{-,(m)})
  \right].
\end{equation}

\subsection{Marginal positivity as a validation criterion}

The framework does not enforce that the marginals $\int f\,\mathrm{d}\bp$ and
$\int f\,\mathrm{d}\bx$ are non-negative.  Physically, both must be
non-negative (they correspond to $|\psi|^2$ and $|\hat\psi|^2$).  Rather than
imposing these constraints architecturally, we use marginal non-negativity as a
post-hoc validation criterion: if the trained solution produces non-negative
marginals, this provides additional evidence that the equation has been solved
correctly.

\section{Adversarial Training Algorithm}\label{sec:algorithm}

\subsection{Loss function}

Substituting the signed estimator~\eqref{eq:mc_estimator} into the
residual~\eqref{eq:residual}, the loss function aggregates over $K$ test
functions:
\begin{equation}\label{eq:loss}
  \mathcal{L}_{\mathrm{total}}[\vt^+,\vt^-,\alpha,\eta]
  = \frac{1}{K}\sum_{k=1}^K \bigl(\hat{R}^{(k)}\bigr)^2,
\end{equation}
where $\hat{R}^{(k)}$ is the Monte Carlo estimate of $R^{(k)}$ using the
signed estimator, and $\eta = \{\bw_x^{(k)},\bw_p^{(k)},\kappa^{(k)},
b^{(k)}\}_{k=1}^K$ collects all test function parameters.

\subsection{Min-max optimization}

The adversarial training objective is:
\begin{equation}\label{eq:minmax}
  \min_{\vt^+,\vt^-,\alpha}\;
  \max_\eta\;
  \mathcal{L}_{\mathrm{total}}[\vt^+,\vt^-,\alpha,\eta].
\end{equation}
The generator (pushforward networks $F^\pm$ and weight $\alpha$) takes
gradient descent steps to minimize $\mathcal{L}_{\mathrm{total}}$; the
adversary (test function parameters $\eta$) takes gradient ascent steps to
maximize it.  This ensures the learned distribution satisfies the weak
formulation against a broad and adaptive set of test functions.

\subsection{Practical implementation}

The training loop follows the standard WANPM
procedure~\cite{He2025,HeCai2026MV}:
\begin{enumerate}[label=\arabic*.]
  \item \textbf{Sample:} Draw $M$ tuples of initial conditions, base noise,
    and time points.  Compute pushforward samples via~\eqref{eq:samples}.
  \item \textbf{Evaluate:} For each test function $k$ and each sample $m$,
    compute the integrand of~\eqref{eq:residual}
    using the signed estimator~\eqref{eq:mc_estimator}.  This requires
    two evaluations of $V$ per sample per test function.
  \item \textbf{Compute loss:} Assemble $\hat{R}^{(k)}$ and
    $\mathcal{L}_{\mathrm{total}}$.
  \item \textbf{Update:} Gradient descent on $(\vt^+,\vt^-,\alpha)$; gradient
    ascent on $\eta$.
\end{enumerate}

The computational cost per sample per test function is: two evaluations of $V$
(at the shifted positions $\bx\pm\frac{\hbar}{2}\bw_p^{(k)}$), one inner
product $\bw_x^{(k)}\cdot\bp$, and one cosine evaluation.  No automatic
differentiation through the potential or through the pushforward network's
Jacobian is required.

\begin{algorithm}[t]
\caption{WANPM for the Wigner--Moyal Equation (one epoch)}
\label{alg:wanpm_wigner}
\begin{algorithmic}[1]
\For{adversary step $= 1,\ldots,n_{\mathrm{adv}}$}
  \State Sample $\{t^{(m)}, \bz_0^{\pm,(m)}, \bz^{\pm,(m)}\}_{m=1}^M$
  \State Compute $(\bx^{\pm,(m)},\bp^{\pm,(m)})$ via \eqref{eq:samples}
  \State Evaluate $\hat{R}^{(k)}$ for $k=1,\ldots,K$ using signed
    estimator \eqref{eq:mc_estimator}
  \State $\eta \gets \eta + \lambda_{\mathrm{adv}}\,
    \nabla_\eta\mathcal{L}_{\mathrm{total}}$
    \Comment{Gradient ascent}
\EndFor
\State Sample new batch; compute $(\bx^{\pm,(m)},\bp^{\pm,(m)})$ and
  $\hat{R}^{(k)}$
\State $(\vt^+,\vt^-,\alpha) \gets (\vt^+,\vt^-,\alpha) -
  \lambda_{\mathrm{gen}}\,
  \nabla_{(\vt^+,\vt^-,\alpha)}\mathcal{L}_{\mathrm{total}}$
  \Comment{Gradient descent}
\end{algorithmic}
\end{algorithm}

\section{Discussion}\label{sec:discussion}

\subsection{Comparison with existing methods}

The method differs from existing approaches to the Wigner equation in several
respects.  Grid-based methods (finite difference, spectral)
discretize the $2N$-dimensional phase space and require $O(n^{2N})$ grid
points, making them infeasible for $N > 2$ or $3$.  The truncated Wigner
approximation~\cite{Steel1998,Polkovnikov2010} replaces $\Theta[V]$ with the
leading terms of its Taylor expansion, introducing systematic errors
proportional to $\hbar^2$.  Signed-particle methods~\cite{Nedjalkov2004,
Sellier2015} sample the phase space stochastically but suffer from the
numerical sign problem.

The present method avoids spatial discretization entirely (the pushforward
network produces samples, not grid values), treats the potential exactly
(no $\hbar$-truncation), and replaces the branching process of signed-particle
methods with a deterministic neural pushforward that can be queried at any
time and any phase-space point.  The signed decomposition~\eqref{eq:decomposition}
is trained to minimize the weak-form residual, rather than being generated by
a stochastic branching rule, which may offer better variance properties.

\subsection{Scaling considerations}

The method inherits the scaling properties of the standard WANPM
framework~\cite{He2025}: the cost per training step is $O(MK)$ evaluations of
$V$ (two per sample per test function), independent of the spatial
dimension~$N$.  The dimension enters only through the size of the pushforward
network (input dimension $1 + 2N + d_{\mathrm{base}}$, output dimension $2N$)
and the number of test function parameters ($2N + 2$ per test function).
For separable potentials $V(\bx) = \sum_i U(x_i)$, the shifted evaluations
$V(\bx\pm\frac{\hbar}{2}\bw_p)$ decompose into $N$ independent scalar
evaluations.

\subsection{Relation to the Weyl quantization}

The identity~\eqref{eq:I_result} has a natural interpretation in terms of the
Weyl correspondence~\cite{Weyl1927}.  The plane-wave test function
$\mathrm{e}^{\ii(\bw_x\cdot\bx + \bw_p\cdot\bp)}$ is the Wigner transform of
the displacement operator $\hat{D}(\bw_x,\bw_p)$, and the
integral $\int(\Theta[V]f)\varphi\,\mathrm{d}\bx\,\mathrm{d}\bp$ computes the
expectation of the corresponding quantum observable.  The reduction to
$V(\bx\pm\frac{\hbar}{2}\bw_p)$ reflects the fact that displacement operators
shift the position argument of the potential, connecting the weak formulation
directly to the Heisenberg picture.

\section{Numerical Experiments}\label{sec:numerics}

We validate the method on two benchmark problems for the one-dimensional
Wigner--Fokker--Planck (WFP) equation, adapted from Yi and
Xu~\cite{YiXu2025}.  The WFP equation augments the Wigner--Moyal
equation~\eqref{eq:wigner} with a Fokker--Planck collision operator modeling
dissipation and decoherence in open quantum systems:
\begin{equation}\label{eq:WFP}
  \frac{\partial W}{\partial t}
  + \xi\,\frac{\partial W}{\partial x}
  + \Theta[V]\,W
  = D_{pp}\frac{\partial^2 W}{\partial\xi^2}
  + 2\gamma\frac{\partial(\xi W)}{\partial\xi}
  + D_{qq}\frac{\partial^2 W}{\partial x^2}
  + 2D_{pq}\frac{\partial^2 W}{\partial x\,\partial\xi},
\end{equation}
where $\varepsilon$ is the semiclassical parameter, $\gamma$ is the friction
coefficient, and $D_{pp}$, $D_{qq}$, $D_{pq}$ are diffusion coefficients.
The collision terms on the right-hand side are standard second-order
differential operators; after integration by parts, they transfer onto the
test function in exactly the same manner as in the classical Fokker--Planck
setting~\cite{He2025}.  The quantum potential term $\Theta[V]W$ is handled
by our central result~\eqref{eq:I_result}, with $\hbar$ replaced by the
semiclassical parameter~$\varepsilon$.

For both experiments, the initial condition is a Gaussian wavepacket
\begin{equation}\label{eq:gaussian_ic}
  W_0(x,\xi)
  = \frac{\sqrt{a_{11}a_{22}-a_{12}^2}}{\pi\varepsilon}
    \exp\!\left[
      -\frac{a_{11}(x-x_0)^2 + a_{22}(\xi-\xi_0)^2 + 2a_{12}(x-x_0)(\xi-\xi_0)}
            {\varepsilon}
    \right]
\end{equation}
with $\varepsilon=0.1$, $(x_0,\xi_0)=(0.1,-0.2)$, $a_{11}=a_{22}=1$,
$a_{12}=0$.  The physical parameters are $\gamma=1$, $D_{pp}=D_{qq}=0.2$,
$D_{pq}=0.05$, and the domain is $[-2,2]\times[-2,2]$ with $T=0.5$.  The
signed pushforward uses $d_{\mathrm{base}}=8$, $K=2000$ test functions,
batch size $M=2000$, and is trained with Adam (learning rate $10^{-3}$).

\subsection{Experiment 1: harmonic potential}

The potential is $V(x) = \tfrac{1}{2}x^2 + x$.  For a quadratic potential
the Wigner operator $\Theta[V]$ reduces to the classical Liouville operator,
so the quantum corrections vanish and the Wigner function remains Gaussian
for all time.  The exact solution is obtained from the closed-form covariance
evolution (method of characteristics for the Gaussian system).  Training uses
$1{,}500$ epochs.

\begin{figure}[t]
\centering
\includegraphics[width=\textwidth]{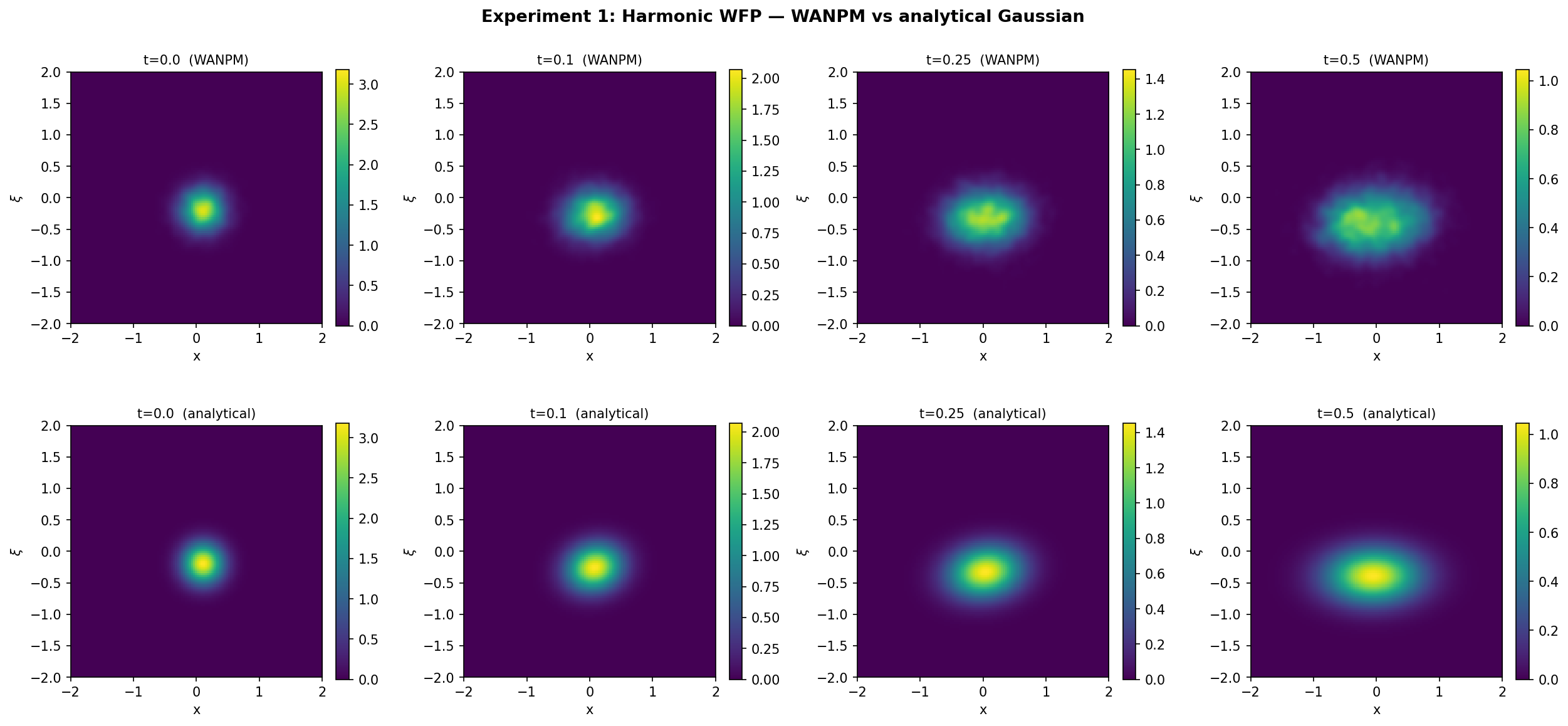}
\caption{Experiment~1 (harmonic $V$): phase-space snapshots of the Wigner
function at $t=0, 0.1, 0.25, 0.5$.  Top: WANPM (KDE from $8{,}000$ samples).
Bottom: analytical Gaussian.}
\label{fig:exp1_snapshots}
\end{figure}

\begin{figure}[t]
\centering
\includegraphics[width=\textwidth]{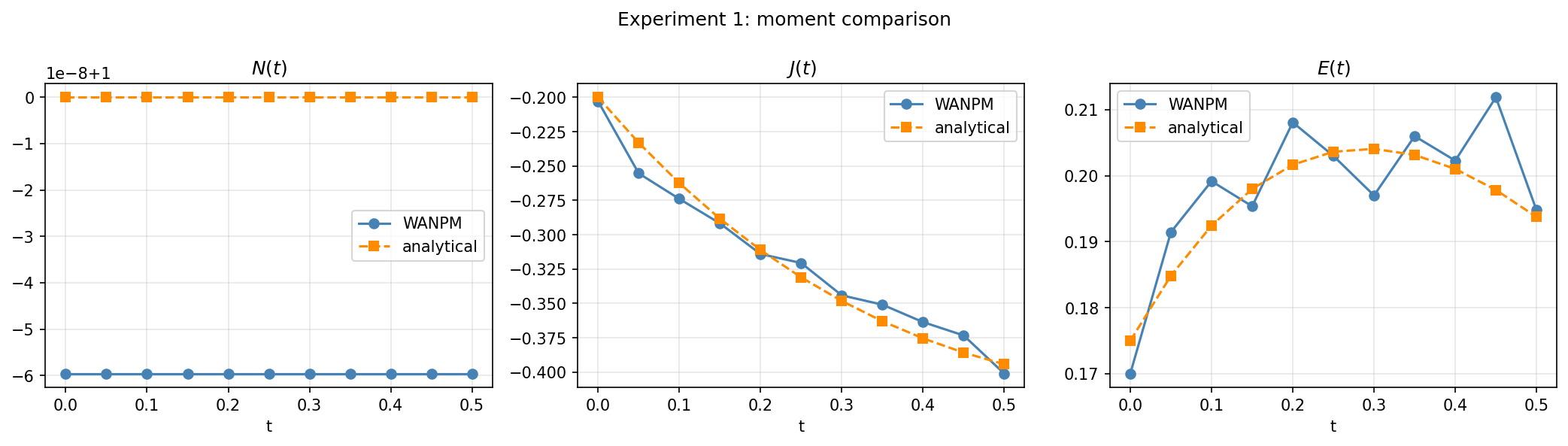}
\caption{Experiment~1: macroscopic moments $N(t)$, $J(t)$, $E(t)$ comparing
WANPM (blue circles) and the analytical reference (orange squares).}
\label{fig:exp1_moments}
\end{figure}

Figure~\ref{fig:exp1_snapshots} compares the learned phase-space density
(obtained via kernel density estimation on pushforward samples) against the
analytical Gaussian at four time snapshots.  The WANPM solution accurately
tracks the spreading and drift of the wavepacket throughout $[0, 0.5]$.
Figure~\ref{fig:exp1_moments} compares the macroscopic moments $N(t)$
(total particle number), $J(t)$ (total momentum), and $E(t)$ (total energy)
between WANPM and the analytical reference.  The momentum $J(t)$ is captured
accurately, and the total particle number $N(t)$ is preserved to within
$O(10^{-8})$.

\begin{figure}[t]
\centering
\includegraphics[width=\textwidth]{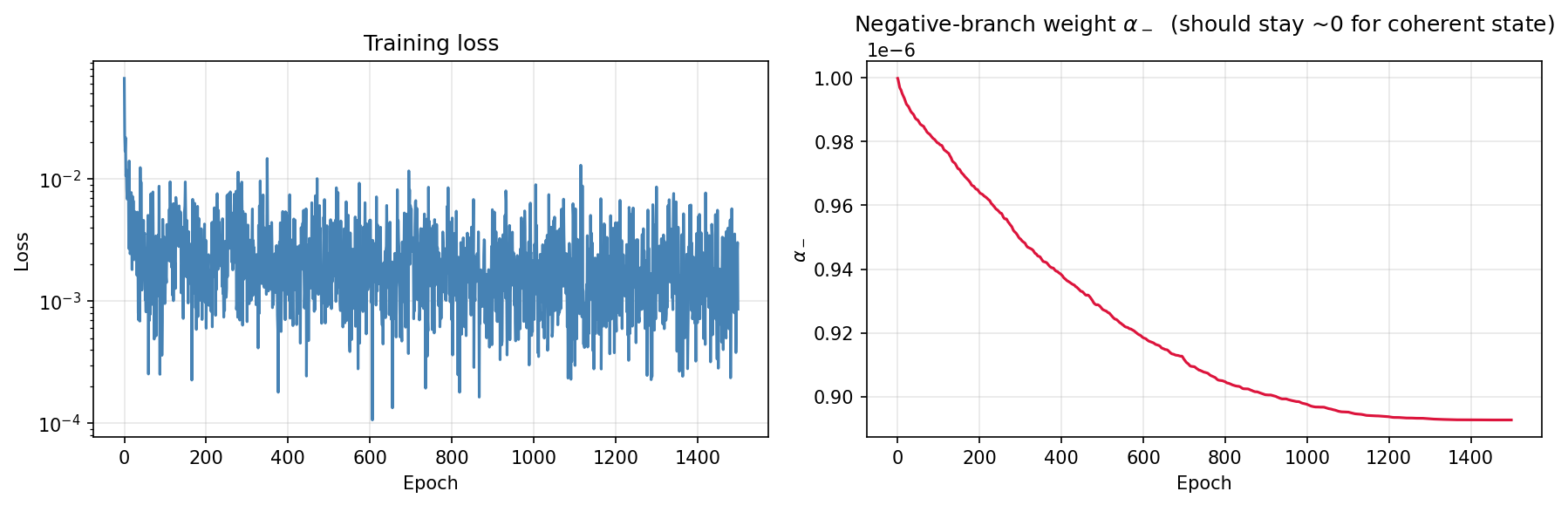}
\caption{Experiment~1: training loss (left) and learnable negative-branch
weight $\alpha^-$ (right).  The weight decays to $O(10^{-6})$, confirming
that the solution remains non-negative.}
\label{fig:exp1_training}
\end{figure}

Figure~\ref{fig:exp1_training} shows the training loss converging to
$O(10^{-3})$ and the learnable weight $\alpha^-$ decaying toward zero during
training --- consistent with the fact that the Wigner function of a coherent
state is everywhere non-negative, so no negative branch is needed.

\subsection{Experiment 2: double-well potential}

The potential is $V(x) = (x^2-1)^2$, a symmetric double well.  This is a
genuinely non-harmonic potential for which the quantum correction
$\Theta[V]W$ does not reduce to a classical operator, and no closed-form
solution is available.  The reference solution is computed using the
time-splitting Fourier pseudospectral (TSSP) method of Yi and
Xu~\cite{YiXu2025} on a $256\times 256$ phase-space grid with
$\Delta t = 1/512$.  Training uses $5{,}000$ epochs.

\begin{figure}[t]
\centering
\includegraphics[width=\textwidth]{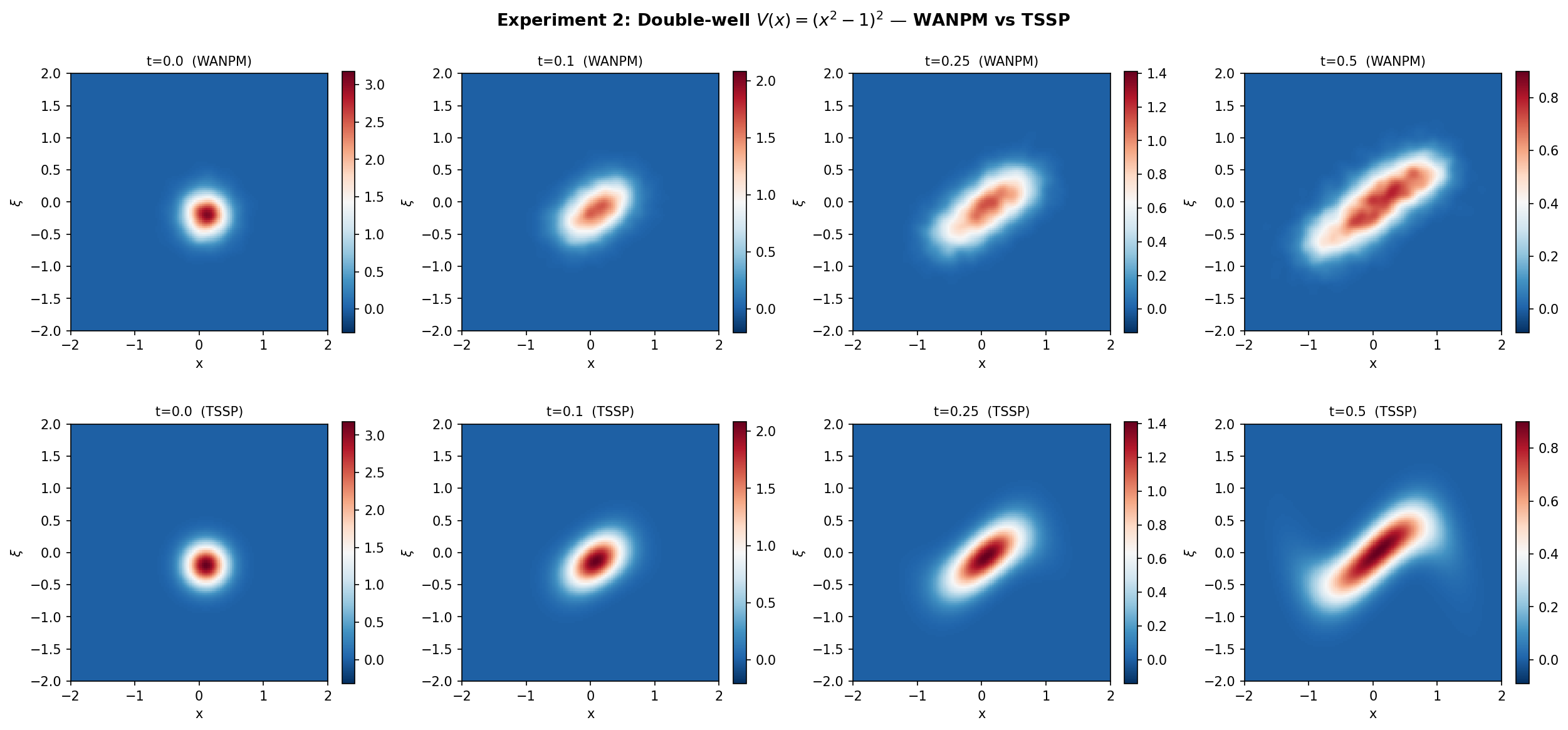}
\caption{Experiment~2 (double-well $V(x)=(x^2-1)^2$): phase-space snapshots
at $t=0, 0.1, 0.25, 0.5$.  Top: WANPM.  Bottom: TSSP reference.}
\label{fig:exp2_snapshots}
\end{figure}

Figure~\ref{fig:exp2_snapshots} compares the WANPM-learned density against
the TSSP reference at four time snapshots.  The initial Gaussian wavepacket
spreads and deforms under the influence of the double-well potential, and the
WANPM solution captures both the overall shape and the asymmetric elongation
visible at $t=0.5$.  The finite-difference treatment of the potential
$[V(x+\varepsilon w_\xi/2) - V(x-\varepsilon w_\xi/2)]/\varepsilon$ handles
the quartic nonlinearity exactly, without any $\varepsilon$-truncation.

\begin{figure}[t]
\centering
\includegraphics[width=\textwidth]{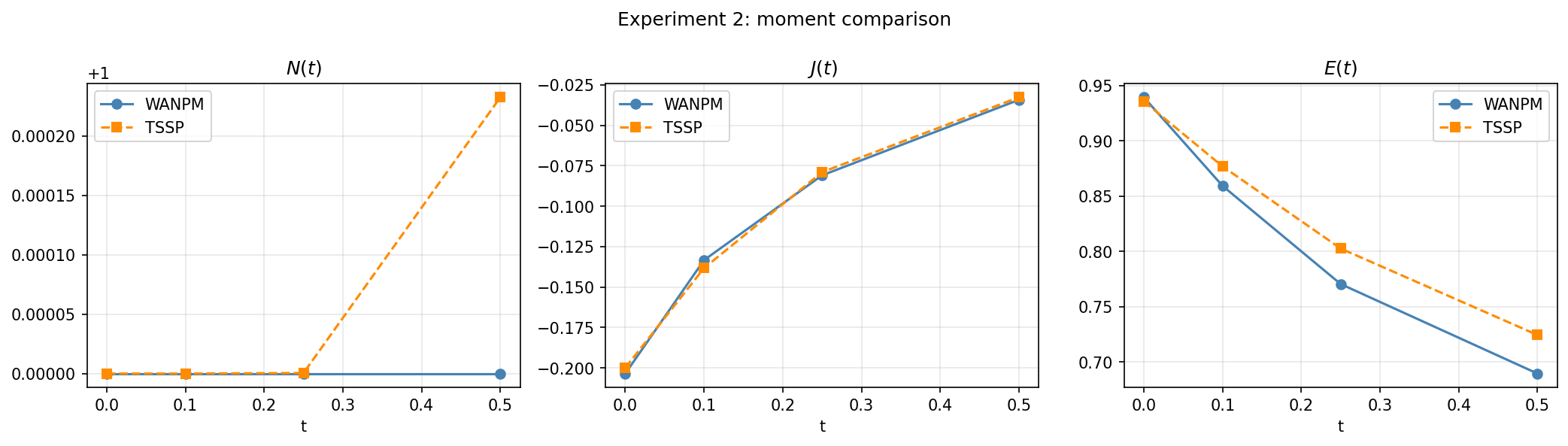}
\caption{Experiment~2: macroscopic moments $N(t)$, $J(t)$, $E(t)$ comparing
WANPM (blue circles) and the TSSP reference (orange squares).}
\label{fig:exp2_moments}
\end{figure}

Figure~\ref{fig:exp2_moments} compares the macroscopic moments.  The total
momentum $J(t)$ shows excellent agreement with the TSSP reference.  The
total energy $E(t)$ shows some deviation at later times, which we attribute
to the finite expressivity of the pushforward network and the short training
duration; longer training and larger networks are expected to improve this.

\begin{figure}[t]
\centering
\includegraphics[width=\textwidth]{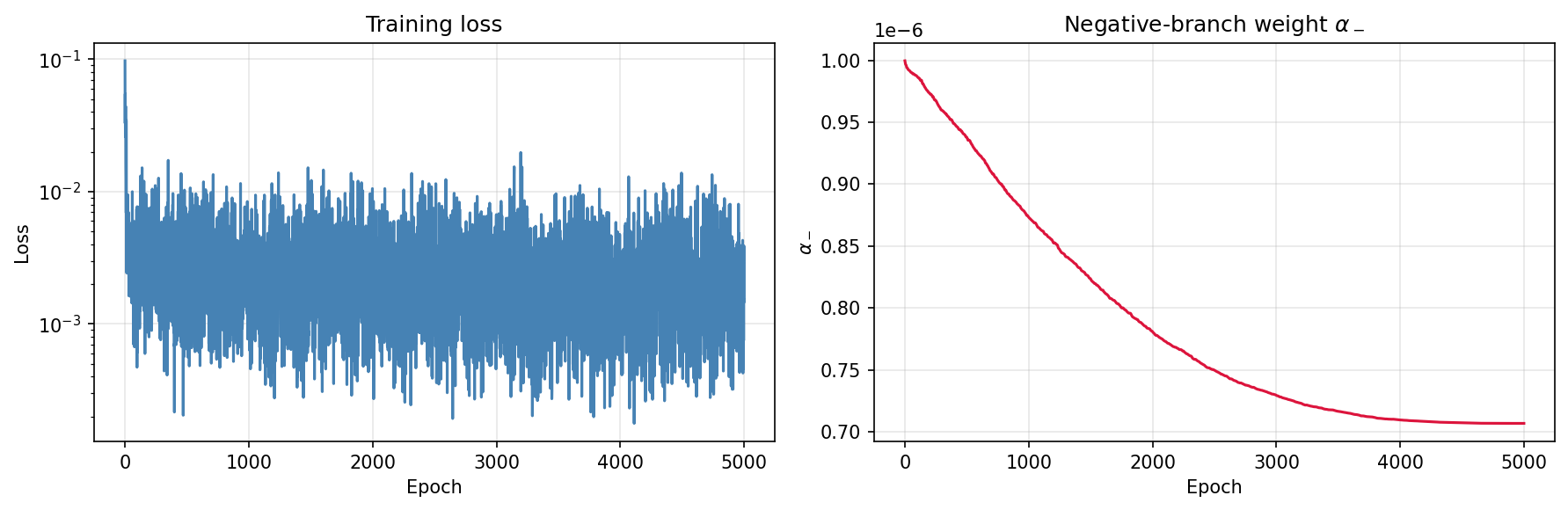}
\caption{Experiment~2: training loss (left) and learnable negative-branch
weight $\alpha^-$ (right).}
\label{fig:exp2_training}
\end{figure}

Figure~\ref{fig:exp2_training} shows the training dynamics.  The loss
stabilizes in the $10^{-3}$ range, and $\alpha^-$ again decays toward zero,
indicating that the Wigner function remains predominantly non-negative for
this initial condition and time horizon.

\section{Conclusion}\label{sec:conclusion}

We have shown that the WANPM framework with plane-wave test functions extends
naturally to the full, untruncated Wigner--Moyal equation.  The Fourier
structure of the plane-wave test function exactly inverts the Fourier transform
defining the Wigner potential kernel, reducing the nonlocal pseudo-differential
operator to a pointwise finite difference of the potential.  This yields a
weak-form residual~\eqref{eq:residual} that requires only two evaluations of
$V$ per sample per test function, with no derivatives and no $\hbar$-expansion.

The signed pushforward architecture~\eqref{eq:decomposition} handles the
negativity of the Wigner function by decomposing it into two non-negative
phase-space distributions with a learnable mixing weight.  Marginal
non-negativity serves as a post-hoc validation criterion rather than an
architectural constraint.

Numerical experiments on the Wigner--Fokker--Planck equation
(Section~\ref{sec:numerics}) validate the method on both a harmonic potential
(where the exact Gaussian solution is recovered) and a double-well potential
(where the WANPM solution agrees with a high-resolution TSSP reference).  The
learnable weight $\alpha^-$ correctly decays to near zero for initial
conditions that remain non-negative.

The practical consequences are: (i)~exact treatment of the quantum potential
for any $\hbar$; (ii)~black-box access to $V$ with no derivative information;
(iii)~the same mesh-free, Jacobian-free, scalable training loop as the
standard WANPM framework; and (iv)~a deterministic signed pushforward that
avoids the variance growth of stochastic signed-particle methods.



\begin{thebibliography}{99}

\bibitem{Wigner1932}
E.~Wigner, On the quantum correction for thermodynamic equilibrium,
Phys.\ Rev.\ 40 (1932) 749--759.

\bibitem{Moyal1949}
J.~E.~Moyal, Quantum mechanics as a statistical theory,
Math.\ Proc.\ Cambridge Philos.\ Soc.\ 45 (1949) 99--124.

\bibitem{Steel1998}
M.~J.~Steel, M.~K.~Olsen, L.~I.~Plimak, P.~D.~Drummond, S.~M.~Tan,
M.~J.~Collett, D.~F.~Walls, R.~Graham,
Dynamical quantum noise in trapped Bose--Einstein condensates,
Phys.\ Rev.\ A 58 (1998) 4824--4835.

\bibitem{Polkovnikov2010}
A.~Polkovnikov,
Phase space representation of quantum dynamics,
Ann.\ Phys.\ 325 (2010) 1790--1852.

\bibitem{Nedjalkov2004}
M.~Nedjalkov, H.~Kosina, S.~Selberherr, C.~Ringhofer, D.~K.~Ferry,
Unified particle approach to Wigner--Boltzmann transport in small semiconductor
devices,
Phys.\ Rev.\ B 70 (2004) 115319.

\bibitem{Sellier2015}
J.~M.~Sellier, I.~Dimov,
The Wigner--Boltzmann Monte Carlo method applied to electron transport in the
presence of a single dopant,
Comput.\ Phys.\ Commun.\ 185 (2014) 2427--2435.

\bibitem{PINN}
M.~Raissi, P.~Perdikaris, G.~E.~Karniadakis,
Physics-informed neural networks: A deep learning framework for solving forward
and inverse problems involving nonlinear partial differential equations,
J.~Comput.\ Phys.\ 378 (2019) 686--707.

\bibitem{deepritz}
W.~E, B.~Yu,
The Deep Ritz method: A deep learning-based numerical algorithm for solving
variational problems,
Commun.\ Math.\ Stat.\ 6 (2018) 1--12.

\bibitem{wan}
Y.~Zang, G.~Bao, X.~Ye, H.~Zhou,
Weak adversarial networks for high-dimensional partial differential equations,
J.~Comput.\ Phys.\ 411 (2020) 109409.

\bibitem{He2025}
A.~Q.~He, W.~Cai,
Neural pushforward samplers for transient distributions from Fokker--Planck
equations with weak adversarial training,
arXiv:2509.14575, 2025.

\bibitem{HeCai2026MV}
A.~Q.~He, W.~Cai,
Weak adversarial neural pushforward method for the McKean--Vlasov / mean-field
Fokker--Planck equation,
arXiv:2603.16186, 2026.

\bibitem{HeCai2026manifold}
A.~Q.~He, W.~Cai,
Weak adversarial neural pushforward method for Fokker--Planck equations on
Riemannian manifolds,
in preparation, 2026.

\bibitem{Weyl1927}
H.~Weyl,
Quantenmechanik und Gruppentheorie,
Z.~Phys.\ 46 (1927) 1--46.

\bibitem{YiXu2025}
Q.~Yi, L.~Xu,
A time-splitting Fourier pseudospectral method for the
Wigner(-Poisson)-Fokker-Planck equations,
arXiv:2509.11153, 2025.

\end{thebibliography}
\end{document}